\documentclass[twocolumn,showpacs,preprintnumbers,amsmath,amssymb]{revtex4}

\usepackage{graphicx}
\usepackage{dcolumn}
\usepackage{bm}

\begin{document}

\preprint{APS/123-QED}

\title{Correlated Mixture Between Adiabatic and Isocurvature\\
Fluctuations and Recent CMB Observations}

\author{Ana Paula A. Andrade}
 \email{apaula@uesc.br}
\affiliation{%
Divis\~ao de Astrof\'{\i}sica, Instituto Nacional de Pesquisas
Espaciais/MCT}%
\affiliation{Laborat\'orio de Astrof\'isica Te\'orica e Observacional\\
DCET/UESC\\}%

\author{Carlos Alexandre Wuensche}%
 \email{alex@das.inpe.br}
\affiliation{%
Divis\~ao de Astrof\'{\i}sica, Instituto Nacional de Pesquisas
Espaciais/MCT}%

\author{Andr\'e Lu\'is Batista Ribeiro}
 \email{albr@uesc.br}
\affiliation{Laborat\'orio de Astrof\'isica Te\'orica e Observacional\\
DCET/UESC\\}%

\date{\today}

\begin{abstract}
This work presents a reduced $\chi^2_\ensuremath{\nu}$ test to
search for non-gaussian signals in the CMBR TT power spectrum of
recent CMBR data, WMAP, ACBAR and CBI data sets,
assuming a mixed density field including adiabatic and
isocurvature fluctuations. We assume a skew positive mixed model
with adiabatic inflation perturbations plus additional
isocurvature perturbations possibly produced by topological
defects. The joint probability distribution used in this context
is a weighted combination of Gaussian and non-Gaussian random
fields. Results from simulations of CMBR temperature 
for the mixed field show a distinct signature in CMB
power spectrum for very small deviations (\ensuremath{\sim} 0.1\%)
from a pure Gaussian field, and can be used as a direct test for
the nature of primordial fluctuations. A reduced
$\chi^2_\ensuremath{\nu}$ test applied on the most recent CMBR
observations reveals that an isocurvature fluctuations field is
not ruled out and indeed permits a very good description for a
flat geometry \ensuremath{\Lambda}-CDM universe, $\chi^2_{930}$\ensuremath{\sim} 1.5,
 rather than the
simple inflationary standard model with
$\chi^2_{930}$\ensuremath{\sim} 2.3. This result may looks is particular discrepant with the 
reduced $\chi^2$ of 1.07 obtained with the same model in Spergel et al. (2003) 
for temperature only, however, our work is restricted to a region of 
the parameter space that does not include the best fit model for TT only 
of Spergel et al. (2003).

\end{abstract}

\maketitle

\section{\label{sec:level1}Introduction}

The new generation of cosmic microwave background radiation (CMBR)
experiments has opened a new era in astrophysics, the precision
cosmology era. Recent observations, especially those of WMAP
(Wilkinson Anisotropy Microwave Probe), ACBAR (Arcminute Cosmology
Bolometer Array Receiver) and CBI (Cosmic Background Imager),
brought a new light to CMBR fluctuations studies in large,
intermediate and small scales. This new generation of experiments
made possible the characterization of the power spectrum of
temperature fluctuations up to the third acoustic peak (Hinshwaw
et al. 2003; Pearson et al. 2002; Kuo et al. 2004). Indeed, the
combination of CMBR and large scale structure (LSS) data allows
cosmologists to constrain cosmological parameters for a given set
of scenarios and also determine the nature of the primordial
fluctuations.  Since CMBR carries the intrinsic statistical
properties of cosmological perturbations, it is considered the
most powerful tool to investigate the nature of cosmic structure.

The most accepted model for structure formation assumes initial
quantum fluctuations created during inflation and amplified by
gravitational effects. The standard inflation model predicts an
adiabatic uncorrelated random field with a nearly flat,
scale-invariant spectrum on scales larger than \ensuremath{\sim}
1-2\ensuremath{^\circ} (Guth 1981; Salopek, Bond \& Bardeen 1981;
Bardeen, Steinhardt \& Turner 1983). Simple inflationary models
also predict the random field follows a nearly-Gaussian
distribution, where only small deviations from Gaussianity are
allowed (e.g. Gangui et al. 1994). In hybrid inflation models
(Battye \& Weller 1998; Battye, Magueijo \& Weller 1999),
structure is formed by a linear combination of
(inflation-produced) adiabatic and (topological defects induced)
isocurvature density fluctuations. The topological defects are
assumed to appear during the phase transition that marks the end
of the inflationary epoch.

Correlated mixed field with adiabatic and isocurvature
fluctuations has been considered by many authors, as Bucher et al.
(2000), Gordon (2001) and Amendola et al (2002). Andrade et al.
(2004) have suggested a new mixed scenario describing a correlated
mixture of two fields, one of them being a dominant Gaussian
adiabatic process to which is added a small contribution of an
isocurvature field. In a previous work, Ribeiro, Wuensche \&
Letelier (2000) used such mixed scenario to probe the galaxy
cluster abundance evolution in the universe and found that even a
very small level of non-Gaussianity in the mixed field may
introduce significant changes in the cluster abundance rate.
Andrade, Wuensche and Ribeiro (2004) showed the effects of mixed
models in the CMBR power spectrum, combining a Gaussian
(adiabatic) field with a second isocurvature field to produce a
positive skewness mixed density fluctuation field. This approach
adopted a scale dependent mixture parameter and a power-law
initial spectrum to simulate the CMBR temperature and polarization
power spectra for a flat \ensuremath{\Lambda}-CDM model,
generating quite a large grid of cosmological parameters
combination. The results show how the shape and amplitude of the
fluctuations in CMBR depend upon such mixed fields and how we can
easily distinguish a standard adiabatic Gaussian field from a
mixed non-Gaussian one and easily quantify the contribution of the
second component.

In this work, we apply a statistical test to both the mixed power
spectrum simulations and recent CMBR data obtained by WMAP
(Spergel et al. 2003), ACBAR (Kuo et al. 2004) and CBI (Pearson et
al. 2002) in order to compare how well the standard and mixed
models describe the most recent observations. We also estimate the
possible contribution of an isocurvature field to the primordial
density fluctuation field in the mixed model scenario.

\section{Recent CMB Observations}

The experiments ACBAR and CBI present the highest sensitivity and
highest signal to noise observations of CMBR temperature
distribution in small angular scales ( \ensuremath{\sim} 4 -5
arcminutes). In scales larger than the above mentioned
(\ensuremath{\sim} 20 arcminutes ) the WMAP satellite made a
set of all-sky maps measuring both the temperature and
polarization anisotropies, opening a new window for cosmological
investigations.

The WMAP observations, combined with LSS observations has made possible to
constrain a precise picture of the cosmos. Using the three-dimensional power spectrum P(k) 
from over 200,000 galaxies in the Sloan Digital Sky Survey (SDSS) in 
combination with WMAP and other data, Tegmark et al. (2003)
show that recent observations are consistent with a flat adiabatic  
\ensuremath{\Lambda}-CDM model. Specifically, the CMB power spectrum
exhibits a first acoustic peak at {\cal l} = 220.1 \ensuremath{\pm} 0.8,
with amplitude of the 74.7 \ensuremath{\pm} 0.5 \ensuremath{\mu}K
; and a second acoustic peak at {\cal l} = 546 \ensuremath{\pm} 10, with
amplitude of 48.8 \ensuremath{\pm} 0.9 \ensuremath{\mu}K, a
picture consistent with inflation predictions (Peiris et al.
2003). The best fit model compared with the WMAP plus ACBAR and
CBI is for a \ensuremath{\Lambda}-CDM Universe with the following
cosmological parameter combination:
\ensuremath{\Omega}$_{tot}$h$^{2}$ = 1.02 \ensuremath{\pm} 0.002;
\ensuremath{h = 0.71^{+0.04}_{-0.03}}; \ensuremath{\Omega_{CDM}h^2
= 0.0224 \pm 0.0009}; \ensuremath{\Omega_m h^2 =
0.135^{+0.008}_{-0.009}}; $n_s (0.05 Mpc^{-1}) = 0.93 \pm 0.03$
and a slope running spectral index with $dn_s/dln(k) = -0.031^
{+0.016}_{-0.018}$ (Spergel et al. 2003). The present CMBR data
provides a strong support for adiabatic field domination.

However, the WMAP data show several peculiarities at various
values of {\cal l}. WMAP observations show that the fluctuations in large
scale present a lower amplitude than the standard inflation model
predicts. The temperature power spectrum is almost 30\% suppressed
on $\sim$ 1 degree angular scales (Hinshaw et al., 2003),
specially those scales related to the quadrupole ({\cal l} = 2) and the
octopole ({\cal l} = 3), when compared with the predictions of the
standard gaussian \ensuremath{\Lambda}-CDM models. Another impressive 
WMAP conclusions is the evidence of an optical depth
to re-scattering by electrons with a value of \ensuremath{\tau} =
0.17 \ensuremath{\pm} 0.04, describing a reionization scenario at
redshift 11 \ensuremath{\leq} z \ensuremath{\leq} 30 (Kogut et al. 2003).

In this paper, we try to reproduce the main feature of the CMBR
power spectrum, in large, intermediate and small scales, invoking just the primordial physics, by
considering a correlated mixed field in combination with six cosmological
parameters: \ensuremath{\Omega}$_{b}$, \ensuremath{\Omega}$_{cdm}$, 
\ensuremath{\Omega}$_{\ensuremath{\Lambda}}$, n, H$_{0}$, the amplitude of the
power spectrum, {\cal A}, and the mixed
coefficient, \ensuremath{\alpha}$_{0}$. Since we are interested in just cosmological principles, no secondary effects are considered in CMB
fluctuations, like reionization models, gravitational lenses nor 
Sunyayev-Zeldovich effects.

\section{The Mixed Model}

\subsection{The Mixed Primordial Power Spectrum}

In the mixed scenario, we suppose that the field has a probability
density function of the form:
\begin{equation}
{\cal P}[\delta_k]~\propto ~(1-\alpha)f_1(\delta_k) + \alpha
f_2(\delta_k)
\end{equation}
The first field will always be the Gaussian component and a
possible effect of the second component is to modify the Gaussian
field to have a positive tail. The parameter $\alpha$ in (1)
allows us to modulate the contribution of each component to the
resultant field. Like the hybrid inflation models (Battye \&
Weller 1998; Battye, Magueijo \& Weller 1999), the mixture models
consider the scenario in which structure is formed by both
adiabatic density fluctuations produced during inflation and
active isocurvature perturbations created by cosmic defects during
a phase transition which marks the end of inflationary epoch.

Nevertheless, the mixed scenario considers a possible correlation
between the adiabatic and the isocurvature fields on the
post-inflation Universe. So, the fluctuations in super-horizon
scales (inflated during the exponential expansion) are strictly
uncorrelated and the mixing effect acts only inside the Hubble
horizon, in sub-degree scales. To allow for this condition and
keep a continuous mixed field, a scale dependent mixture
parameter, \ensuremath{\alpha} \ensuremath{\equiv}
\ensuremath{\alpha}(k) was defined. We assume the simplest choice of
\ensuremath{\alpha}(k), a linear function of $k$:
\begin{equation}
\alpha(k) \equiv \alpha_{0}k
\end{equation}

To constrain the two component random field, we take $\delta_{k}=
P(k)\nu^2$, where $\nu$ is a random number with
distribution given by (1). So, we consider the primordial power spectrum
of the mixed field in the form:
\begin{equation}
P(k)^{mix} \equiv M^{mix} (\alpha_0,k)P(k)
\end{equation}

\noindent where the $P(k)$ represents a primordial conventional power-law spectrum and
M$^{mix}$($\alpha$) is the mixture term, a functional of
$f_{\mathit{1}}$ and $f_{\mathit{2}}$, which account for the
statistics effect of a new component:
\begin{equation}
M^{mix}(\alpha_0,k)\equiv\int_\nu[(1-\alpha_0k)f_1(\nu)+\alpha_0k
f_2(\nu)]\nu^2\; d\nu
\end{equation}

\noindent The effective mixed primordial power spectrum is expressed as:

\begin{eqnarray}
P(k)^{mix}= k^{n} + M(\alpha_{0})k^{n+1}
\end{eqnarray}

\noindent where M(\ensuremath{\alpha}$_{0}$) represents only the
coefficient dependence, \ensuremath{\alpha}$_{0}$. In the case of
a pure Gaussian field, M(\ensuremath{\alpha}$_{0}$) \ensuremath{\approx} \ensuremath{\alpha}$_{0}$
\ensuremath{\approx} 0, the mixed power spectrum will be a
simple power-law spectrum, $k^{n}$, where n is the conventional spectral index 
predicted by inflationary models. In the case of a mixed field, the phase
correlations between both fields are estimated by the integral in
Eq.(4), on mixture scales defined by Eq. (2).

\subsection{The Evolution of The Mixed Field}

To estimate the CMBR anisotropy we need to evaluate the evolution of
fluctuations field generated in the early universe through the
radiation-dominated era and recombination. In the mixed model, the 
evolution of both adiabatic and isocurvature components of the
mixed density field is considered an
independent process, when only their effective amplitude correlation is
considered on the decoupled surface of the CMBR photons. 

To compute the independent evolution of the adiabatic \ensuremath{\Lambda}-CDM mode and 
the isocurvature \ensuremath{\Lambda}-CDM mode, we have used the most precision code, 
the Linger function of the COSMICS code package (Bertschinger
1999; Ma \& Bertschinger 1995). Linger does generate the most accurate results for
the photon density field, being able to compute the CMBR anisotropy with integrations
errors less than 0.15\%. However, Linger routine is very time consuming, but
for investigation of small deviations from gaussianity, the COSMICS package seems to
be the most precise code for mixed CMB simulations.

Once estimated the photon density evolution, the multipole moments, C$_{l}$, of the CMBR temperature power
spectra were estimated by a mixed photon density function
incorporated to the original COSMICS package:

\begin{equation}
C_l = 4\pi \int_0^{k_{max}}\;d^3k P^{mix}(k)
(\Delta_l^{mix})^2(k,\tau)
\end{equation}

\noindent The function $\Delta_l^{mix}$ represents the mixed
photon density field in the last scattering surface defined by:

\begin{equation}
\Delta_l^{mix}\equiv(1-\alpha_0k)\Delta_l^{Adi} +
\alpha_0k\Delta_l^{Iso}
\end{equation}

\noindent where $\Delta_l^{Adi}$ and $\Delta_l^{Iso}$ are the
photon density function estimated by COSMICS, respectively, for an
adiabatic and an isocurvature seed initial conditions.

Inserting (7) in (6), we obtain a mixed term in the C$_{l}$
estimation. This condition suggests that the amplitude of both
fields are cross-correlated at the last scattering surface, with a
mixing ratio defined by \ensuremath{\alpha}$_{0}$ and in a
characteristic scale defined by
\ensuremath{\alpha}$_{0}$\textit{k}. The power spectra estimated
by Eq.(3) consider a flat \ensuremath{\Lambda}-CDM universe
slightly distorted by a non-gaussian statistics with constant
spectral index.

A number of realizations of CMBR temperature power spectra were
made and the mean temperature fluctuations combining Gaussian,
Exponential, Lognormal, Rayleigh, Maxwellian and a Chi-squared
distributions was estimated (Andrade et al. 2004). The simulations
show that the influence of the specific statistics of the second
component in the mixed field is not so important as the
cross-correlation between the amplitudes of both fields. However, 
some important results were obtained. It is
possible to directly assess and quantify the mixture of a
correlated adiabatic and isocurvature non-Gaussian field. This
behavior points to another possibility to extract information
from a CMBR power spectrum: the possibility of detecting weakly
mixed density fields, even if we can not exactly identify the
mixture components distribution.

\begin{figure}
\includegraphics{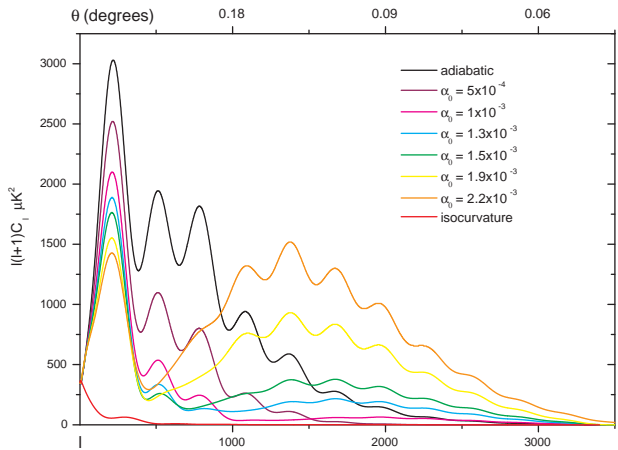}
\caption{\label{fig:epsart} CMBR temperature mixed angular power
spectrum estimated for a \ensuremath{\Lambda}-CDM model in
different mixing degrees.}
\end{figure}

In Figure 1, we see how the shape and amplitude of the spectrum
change even for small values of \ensuremath{\alpha}$_{o}$
(\ensuremath{\sim}10$^{-4}$-10$^{-3}$). The peaks intensities are
clearly susceptible to the existence of mixed fields, although
distinguishing peaks of higher order (2$^{nd}$, 3$^{rd}$, etc.) in
the mixed context is not a straightforward task, since their
intensities, compared to the first peak, are very low. The effect
of increasing \ensuremath{\alpha}$_{0}$ is a power transfer to
smaller scales ({\cal l} \texttt{>}1000), while the super-degree scales
are less affected. However, for large values of
\ensuremath{\alpha}$_{0}$
(\ensuremath{\alpha}$_{0}$ \texttt{>}3x10$^{-3}$) there is a fast
increase in the temperature fluctuations, probably caused by
correlation excess between the mixed fields, resulting in more
power in small scales. Therefore, an acceptable range for
$\alpha_0$ is set to be $\alpha_0 \lesssim 3\times 10^{-3}$.

\section{A $\chi^2$ Test on CMB Observation}

In order to quantify a possible non-gaussian isocurvature
contribution to the fluctuations field, we applied a maximum
likelihood approach to the power spectrum estimated by the most
recent CMBR observations. We ran a set of CMBR realizations for
mixed and pure density fields, considering a flat
\ensuremath{\Lambda}-CDM Universe consistent with inflation predictions and
LSS observations. 

Instead of the usual CMBR analysis approach, which considers
tensors contribution and reionization effects, we consider only
the main six cosmological parameters (\ensuremath{\omega_{cdm}}, \ensuremath{\omega_b}, 
\ensuremath{\omega}$_{\ensuremath{\Lambda}}$, n, 
H$_{0}$, \ensuremath{\alpha_0}), and also {\cal A}, described in the range of
possibilities: 0.8\texttt{<} n \texttt{<}1.2; 0.015
\texttt{<}\ensuremath{\Omega}$_{b}$\texttt{<} 0.03; 0.6
\texttt{<}\ensuremath{\Omega}$_{\ensuremath{\Lambda}}$\texttt{<}
0.8; 60 \texttt{<}H$_{0}$\texttt{<} 80; cold dark matter
density set as 1-(\ensuremath{\Omega}$_{b}$ +
\ensuremath{\Omega}$_{\ensuremath{\Lambda}}$), ranging of 0.170
\texttt{<} \ensuremath{\Omega}$_{cdm}$\texttt{<} 0.385; and the range values of
\ensuremath{\alpha}$_{0}$ set in 0.0 \texttt{<}
\ensuremath{\alpha}$_{0}$ \texttt{<} 3x10$^{-3}$. With this considerations, 
we ensure the simulations are in agreement with
the basic preferences of standard inflation scenarios: flat
geometry (\ensuremath{\Omega}$_{tot}$ \ensuremath{\sim} 1) and a
nearly scale invariant primeval spectrum (n \ensuremath{\sim} 1).
\ensuremath{\Omega}$_{b}$ =(0.019 \ensuremath{\pm} 0.01)h$^{-2}$
is consistent with the mass density of baryons determined by Big
Bang nucleossynthesis and the large scale structure observations,
which suggests that the Hubble constant H$_{0}$ assumes values in
the range (66 \texttt{<} H$_{0}$ \texttt{<} 75) km/sec.Mpc; and a
positive cosmological constant value in the range (0.065\texttt{<}
\ensuremath{\Omega}$_{\ensuremath{\Lambda}}$ \texttt{<}0.75), 
as estimated recently by Tegmark et al. (2003). The amplitude of the primordial
power spectrum, {\cal A}, was estimated by the best fit value for each model simulated.

We applied a reduced $\chi^2$ test (Bevington 1969) between the
CMB data and the power spectra simulations in the mixed context in
four steps. First, we consider the WMAP, CBI and ACBAR data
independently, estimate the reduced $\chi^2$ for all three data
sets and then add all data to obtain a new estimation of the reduced $\chi^2$.
So we estimate the
residual difference for C$_{l}$ estimation in
\ensuremath{\mu}K$^2$, and consider the power spectrum data points
(899 for WMAP; 24 for CBI and 14 for ACBAR) minus 7 degree of
freedom. We create a grid of nearly one hundred simulated values of 
\ensuremath{\alpha}$_{0}$ for each combination of standard cosmological parameters.
Since COSMICS is very time consuming, the variations in the standard cosmological parameters 
were set in order to give just the main direction for the $\chi^2$ face 
variations in cosmological parameters. The most precise models, as defined by
the $\chi^2$ test, are summarized on Table 1, which also contains the best
\ensuremath{\alpha}$_{0}$ value estimated for each parameter combination listed with 68\%
confidence limit.

\begin{table*}
\begin{ruledtabular}
\begin{tabular}{cccccccccc|c}
$H_0$&$\Omega_b$&$\Omega_\Lambda$&n&2\ensuremath{\pi}{\cal A}&$\alpha_{0}$&WMAP&CBI&ACBAR&All
data  &   $Best \alpha_{0}$\\
(km$s^{-1}Mpc^{-1}$) & & & & (\ensuremath{\mu}$K^2$) & & $\chi^2_{892}$ & $\chi^2_{24}$&
$\chi^2_{7}$ & $\chi^2_{930}$  & $\Delta\alpha_{0}$\\ \hline

70 & 0.015 & 0.8 & 1.1 & 92.462 & 0.0 & 2.820 & 1.121 & 2.925 & 2.754 & 0.00066 \\
70 & 0.015 & 0.8 & 1.1 & 117.558 & 5x$10^{-4}$ & 1.590 & 2.999 & 7.259 &
1.677 & \ensuremath{\pm} 0.00045 \\\hline

70 & 0.015 & 0.7 & 1.1 & 113.326 & 0.0 & 3.532 & 1.480 & 3.372 & 3.451 & 0.00061 \\
70 & 0.015 & 0.7 & 1.1 & 152.947 & 5x$10^{-4}$ & 1.779 & 3.608 & 9.236 &
1.888 & \ensuremath{\pm} 0.00036\\\hline

70 & 0.015 & 0.6 & 1.1 & 130.444 & 0.0 & 4.324 & 1.934 & 5.485 & 4.234  & 0.00056  \\
70 & 0.015 & 0.6 & 1.1 & 183.351 & 5x$10^{-4}$ & 2.703 & 4.288 & 11.314 &
2.387 & \ensuremath{\pm} 0.00033\\\hline

70 & 0.023 & 0.8 & 1.1 & 87.918 & 0.0 & 2.422  & 1.000  & 2.796 & 2.379 & 0.00057 \\
70 & 0.023 & 0.8 & 1.1 & 110.492 & 5x$10^{-4}$ & 1.474  & 2.543  & 5.269 &
1.543 & \ensuremath{\pm} 0.00048\\\hline

70 & 0.023 & 0.7 & 1.1 & 109.225 & 0.0 & 2.899 & 1.117 & 2.982 & 2.857   & 0.00052 \\
70 & 0.023 & 0.7 & 1.1 & 145.041 & 5x$10^{-4}$ & 1.560 & 2.875 & 6.959 &
1.651 & \ensuremath{\pm} 0.00034\\\hline

70 & 0.023 & 0.6 & 1.1 & 126.660 & 0.0 & 3.573 & 1.528 & 4.632 & 3.539  & 0.00047 \\
70 & 0.023 & 0.6 & 1.1 & 175.405 & 5x$10^{-4}$ & 2.031 & 3.602 & 9.144 &
2.132 & \ensuremath{\pm} 0.00030\\\hline

70 & 0.030 & 0.8 & 1.1 & 84.160 & 0.0 & 2.124 & 1.000\footnotemark[2] & 3.336 & 2.103  & 0.00047  \\
70 & 0.030 & 0.8 & 1.1 & 104.347 & 5x$10^{-4}$ & 1.490 & 2.270 & 4.663 &
1.555 & \ensuremath{\pm} 0.00036\\\hline

70 & 0.030 & 0.7 & 1.1 & 106.037 & 0.0 & 2.274 & 1.009 & 3.001 & 2.269   & 0.00042 \\
70 & 0.030 & 0.7 & 1.1 & 137.782 & 5x$10^{-4}$ & 1.433 & 2.538 & 5.938 &
1.522\footnotemark[6] & \ensuremath{\pm} 0.00030\\\hline

70 & 0.030 & 0.6 & 1.1 & 123.570 & 0.0 & 2.756 & 1.404 & 4.299 & 2.776  & 0.00038 \\
70 & 0.030 & 0.6 & 1.1 & 166.921 & 5x$10^{-4}$ & 1.855 & 3.146 & 7.950 &
1.956 & \ensuremath{\pm} 0.00029\\\hline

60 & 0.030 & 0.7 & 1.1 & 92.088 & 0.0 & 2.697 & 1.001 & 3.371 & 2.657  & 0.00063 \\
60 & 0.030 & 0.7 & 1.1 & 118.071 & 5x$10^{-4}$ & 1.590 & 2.463 & 5.355 &
1.646 & \ensuremath{\pm} 0.00046\\\hline

65 & 0.030 & 0.7 & 1.1 & 99.315 & 0.0 & 2.410 & 1.001 & 2.791 & 2.383  & 0.00052 \\
65 & 0.030 & 0.7 & 1.1 & 128.530 & 5x$10^{-4}$ & 1.382\footnotemark[1] &
2.344 & 5.398 & 1.456\footnotemark[4] & \ensuremath{\pm} 0.00036\\\hline

75 & 0.030 & 0.7 & 1.1 & 112.147 & 0.0 & 2.212 & 1.243 & 3.707 & 2.231  & 0.00033 \\
75 & 0.030 & 0.7 & 1.1 & 145.624 & 5x$10^{-4}$ & 1.680 & 2.847 & 6.784 &
1.780 & \ensuremath{\pm} 0.00030\\\hline

80 & 0.030 & 0.7 & 1.1 & 117.683 & 0.0 & 2.192 & 1.492 & 4.894 & 2.242  & 0.00026 \\
80 & 0.030 & 0.7 & 1.1 & 151.419 & 5x$10^{-4}$ & 2.081 & 3.228 & 8.150 &
2.190 & \ensuremath{\pm} 0.00030\\\hline

70 & 0.030 & 0.7 & 1.0 & 73.291 & 0.0 & 1.887 & 1.002 & 2.463\footnotemark[3] & 1.863  & 0.00027\\
70 & 0.030 & 0.7 & 1.0 & 92.857 & 5x$10^{-4}$ & 1.668 & 2.855 & 6.798 &
1.778 & \ensuremath{\pm} 0.00034\\\hline

70 & 0.030 & 0.7 & 1.2 & 151.783 & 0.0 & 2.886 & 1.008 & 3.750 & 2.917  & 0.00056 \\
70 & 0.030 & 0.7 & 1.2 & 202.597 & 5x$10^{-4}$ & 1.425 & 2.261 & 5.267 &
1.490\footnotemark[5] & \ensuremath{\pm} 0.00037\\

\end{tabular}
\end{ruledtabular}
\footnotetext[1]{Best model estimated for WMAP data.}
\footnotetext[2]{Best model estimated for CBI data.}
\footnotetext[3]{Best model estimated for ACBAR data.}
\footnotetext[4]{Best model estimated for all data (plot on Figure
2A ).}
\footnotetext[5]{Second model estimated for all data (plot
on Figure 2C).}
\footnotetext[6]{Third model estimated for all data
(plot on Figure 2B).}
\caption{\small The reduced $\chi^2$ test
applied on the WMAP, CBI and ACBAR data and the power spectrum
simulations in a mixed scenario for some \ensuremath{\Lambda}-CDM
models. The last column shows the best \ensuremath{\alpha}$_{0}$ estimation for
each model with 68\% confidence limit}
\end{table*}

The simulations clearly show that WMAP data tends to favor mixed
models. The $\chi^2_\ensuremath{\nu}$ estimated for the fit
between mixed models and WMAP data are significatively lower for
mixing degrees of order \ensuremath{\sim}$10^{-3}$, in all
combination of cosmological parameters tested. However, CBI and
ACBAR data in small angular scales tends to favor the adiabatic
standard models. The combination of all data are more sensible to
WMAP influence (more signal in large and intermediate scales with
low error bars) resulting in best fits models in a mixed scenario.
All data favor a \ensuremath{\Lambda}-CDM Universe with contents
of barions, dark matter and dark energy coherent with that
estimated by the WMAP and LSS analysis.

Figure 2 shows the powers spectrum of the standard
adiabatic and mixed models for the three best fit models obtained
by the $\chi^2_\ensuremath{\nu}$ test. It also shows the behavour of the all
data $\chi^2_\ensuremath{\nu}$ while varying the \ensuremath{\alpha}$_{0}$
value. It is clearly seen that the mixed model adjust the data better than the
standard model in basic combination of cosmological parameters. In this
figure, is possible to observe that the relation between the amplitudes of the
acoustic peaks in the mixed scenario is more adequate to describe
the CMBR power spectrum feature without consideration of any model
of reionization or tensors contributions. Although the mixed model
predicts higher quadrupole and octopole contributions that
predicts the standard adiabatic \ensuremath{\Lambda}-CDM model, the plots clearly show
that the mixed predictions are consistente with acoustic peaks amplitude
estimated by WMAP. Indeed, a model estimation in a good fit is obtained for a
strictly flat Universe, rather than an open Universe or, instead,
a slope running spectral index (Kosowsky \& Turner, 1995) as suggested
by the WMAP team.

One possible explanation of the low amplitude in large scales and the high
amplitude of the secondary acoustic peaks predicted by
the standard model (pure adiabatic) simulated in this work is the lack of the high optical depth, as
considered by the WMAP team (Hinshaw et al. 2003; Spergel et al.
2003). Reionization affects the CMBR in several manners, through Thomson
scattering of photons from free electrons in the intergalactic
medium, which suppress the primordial anisotropies of the CMBR
imprinted on the decoupled surface. In a reionizing scenario, the
acoustic peaks in the CMBR power spectrum would have the amplitude
lowered by $e^{-2\tau}$ and the rescattering would generate large
angle temperature fluctuations through the Doppler effect (Hu \& White,
1997). In this work, rather than considering
reionization effects and invoking just primordial physics, we 
can explain the main feature of the CMBR power spectrum in large, intermediate
and small scales in the context of mixed density field.

\begin{figure*}
\includegraphics[width=5.9in, height=8.0in]{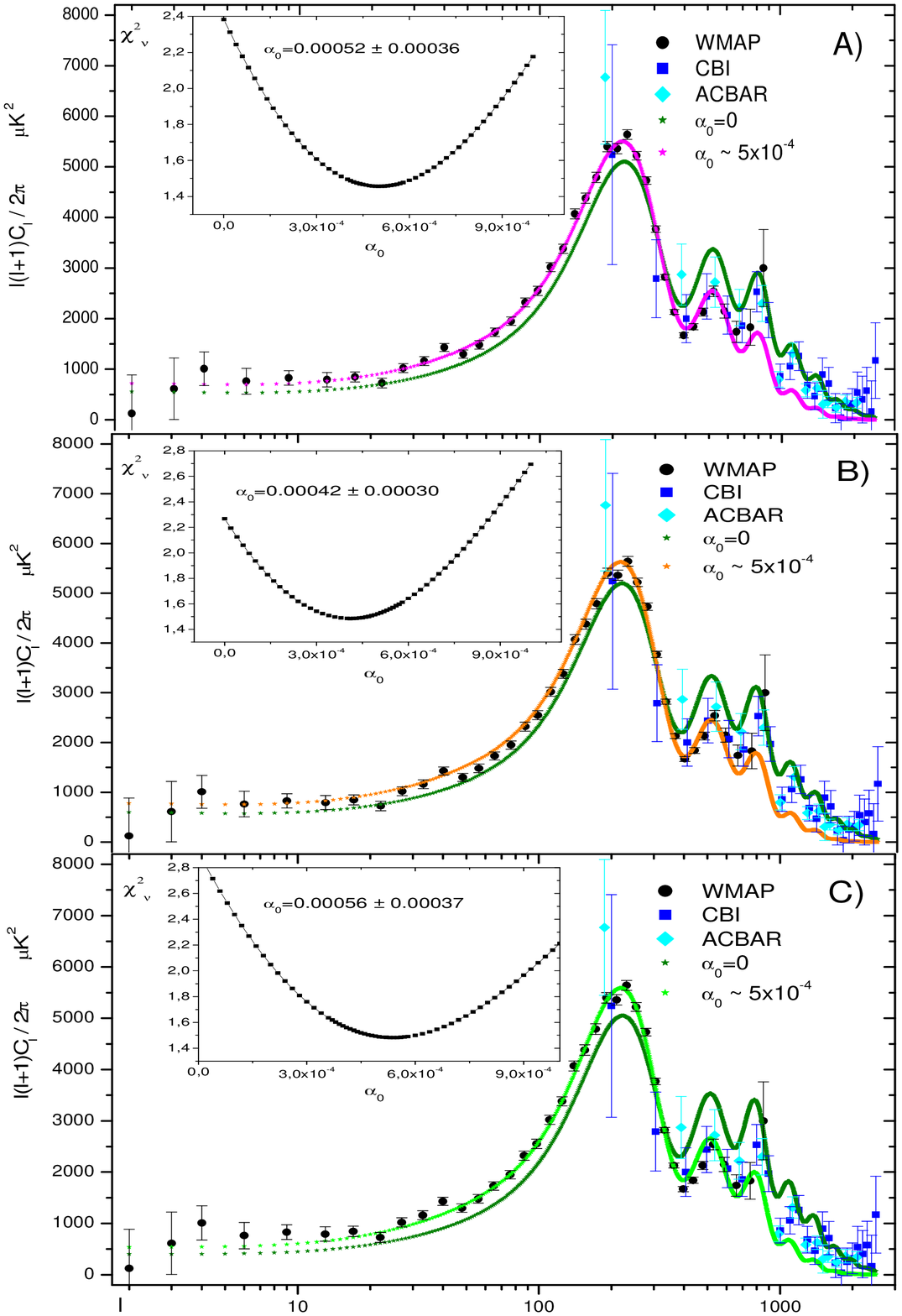}
\caption{\label{fig:epsart} CMBR temperature power spectrum
estimated for the best three parameter combination: \ensuremath{\Omega}$_{b}$= 0.03;
\ensuremath{\Omega}$_{CDM}$= 0.27;
\ensuremath{\Omega}$_{\ensuremath{\Lambda}}$= 0.7. In A: H$_{0}$= 65Km/sMpc and
n=1.1; in B: H$_{0}$= 70Km/sMpc and n=1.2; in C:
H$_{0}$= 70Km/sMpc and
n=1.1. The small picture in figures show the behavior of
$\chi^2_\ensuremath{\nu}$ while varying
\ensuremath{\alpha}$_{0}$ value}
\end{figure*}

\section{Discussion}

We applied a modified statistical $\chi^2_\ensuremath{\nu}$ test
to the most recent CMBR observations. The results allow us to
compare the performance of power spectra simulated with mixed and
standard density fields to describe the mean feature of CMBR
temperature fluctuations. The results showed that the WMAP data
tends to favor mixed density fields with isocurvature fluctuation
contribution of order \ensuremath{\sim}$10^{-3}$ ($\chi^2_{892}$\ensuremath{\sim} 1.5 for mixed model against
$\chi^2_{892}$\ensuremath{\sim} 2.4 in standard models) without the need
of considering any model of reionization. Also, large scale
fluctuations are in good agreement with WMAP measurements. Our result may looks discrepant with the 
reduced $\chi^2$ of 1.07 obtained with the same model in Spergel et al. (2003) 
for temperature only, however, our work is restricted to a region of 
the parameter space that does not include the best fit model for TT only 
of the WMAP team. We can conclude that the mixed scenario offers a good alternative to describe the shape of
TT CMBR power spectrum. Indeed, the data favors a \ensuremath{\Lambda}-CDM Universe with barions, dark
matter and dark energy contents which are coherent with a strictly
flat Universe, with no need to include tensor contributions nor a
slope running spectral index. For a \ensuremath{\Lambda}-CDM Universe with contents of 
\ensuremath{\Omega}$_{b}$= 0.03; 
and the best value indicated for the spectral index, by the $\chi^2$ test, 
n\ensuremath{\sim}1.1; the contribution of the isocurvature field is estimated as  
\ensuremath{\alpha}$_{0}$= 0.00042 \ensuremath{\pm} 0.0003 with 68\% confidence limit.

Although the standard \ensuremath{\Lambda}-CDM model fits well a small 
number of data points of CBI ($\chi^2_{24}$\ensuremath{\sim} 1 for standard
models against
$\chi^2_{24}$\ensuremath{\sim} 2.3 in mixed models) and ACBAR
($\chi^2_7$\ensuremath{\sim} 2.4 for standard models against
$\chi^2_7$\ensuremath{\sim} 6.8 in mixed models), when considering all data points,
CBI, ACBAR and WMAP, the mixed model clearly shows ability to
describe the mean feature of TT CMBR
observation in large, intermediate and small scales. With the upcoming generation of experiments
which will probe temperature fluctuations in small scales, such as
Planck satellite, it will be possible to set new, more stringent,
estimates of the parameter spaces and isocurvature fluctuations contributions. Also, we will try to
constrain the mixed model predictions upon the temperature-polarization
cross correlated power spectrum, in order to consider reinonizations models 
in the mixed scenario. Presently, we conclude that isocurvature fluctuations can not be
ruled out in cosmological studies. Also, conflicting with the
results of WMAP map team (Komatsu et al. 2003), the results of
some independent searches for non-gaussianity in WMAP maps
evidence that WMAP data are consistent with gaussian condition for
{\cal l} \texttt{<}250, marginally gaussian for 224\texttt{<}{\cal l}
\texttt{<}350, and non-gaussian for {\cal l} \texttt{>}350, as revealed,
for instance, by the phase mapping technique (Chiang et al. 2003;
Coles et al. 2003). This picture points an
extra expectation in the study of the mixed model predictions, altough, in
order to constrain a better estimative of \ensuremath{\alpha}$_{0}$,
reionizations models must be investigated in the context of mixed fields. 

The next step in the investigation of the
mixed models in CMBR features will be the predictions of 
the reionization effects, even in a low redshifts, the degeneraices between 
cosmological parameters in mixed context, the predictions of
the angular correlation function and the temperature-polarization
cross correlated power spectrum. 

\begin{acknowledgments}
The authors acknowledge Edmund Bertschinger for the use of the
COSMICS package, funded by NSF under grant AST-9318185. 
APAA thanks the financial support of CNPq and FAPESB, 
under grant 1431030005400. 
CAW was partially supported by CNPq grant 300409/97-4. ALBR thanks 
the financial support of CNPq, under grant 470185/2003-1.

\end{acknowledgments}


\end{document}